\documentclass[twocolumn, twoside, 11pt]{IEEEtran}

\usepackage{times,fancyhdr}
\usepackage{array}

\usepackage{amsmath,amsfonts,amssymb,graphicx,subfigure,epsfig,enumerate,bm,color,float}
\usepackage{ccaption}

\def \0{\mathbf{0}}

\title{Coordinating Complementary Waveforms for Sidelobe Suppression}
\author{{Wenbing Dang,$^{1}$ Ali Pezeshki,$^{1}$ Stephen Howard,$^{2}$ William Moran,$^3$ and Robert Calderbank$^4$}\\
\authorblockA{$^{1}$Colorado State University, Fort Collins, CO, USA\\
$^{2}$Defence Science and Technology Organisation, Edinburgh, SA, Australia\\
$^{3}$University of Melbourne, Melbourne, VIC, Australia\\
$^{4}$Duke University, Durham, NC, USA}\thanks{This work is supported by NSF under grants CCF-0916314 and CCF-1018472 and by DARPA under contract N66001-11-C-4023.}
}

\begin{document}
\maketitle

\begin{abstract}
We present a general method for constructing radar transmit pulse trains and receive filters for which the radar point-spread function in delay and Doppler, given by the cross-ambiguity function of the transmit pulse train and the pulse train used in the receive filter, is essentially free of range sidelobes inside a Doppler interval around the zero-Doppler axis. The transmit pulse train is constructed by coordinating the transmission of a pair of Golay complementary waveforms across time according to zeros and ones in a binary sequence $\mathcal{P}$. The pulse train used to filter the received signal is constructed in a similar way, in terms of sequencing the Golay waveforms, but each waveform in the pulse train is weighted  by an element from another sequence $\mathcal{Q}$. We show that a spectrum jointly determined by $\mathcal{P}$ and $\mathcal{Q}$ sequences controls the size of the range sidelobes of the cross-ambiguity function and by properly choosing $\mathcal{P}$ and $\mathcal{Q}$ we can clear out the range sidelobes inside a Doppler interval around the zero-Doppler axis. The joint design of $\mathcal{P}$ and $\mathcal{Q}$ enables a trade-off between the order of the spectral null for range sidelobe suppression and the signal-to-noise ratio at the receiver output. We establish this trade-off and derive a necessary and sufficient condition for the construction of $\mathcal{P}$ and $\mathcal{Q}$ sequences that produce a null of a desired order.
\end{abstract}

\section{Introduction}

Modern radars are increasingly being equipped with arbitrary waveform generators which enable generation of different wavefields across space, time, frequency, polarization, and wavenumber; see, e.g.,
\cite{Cochran-SPM09}--\nocite{Calderbank-SPM09,Rabi2,Bliss1,Fried4,Nehorai,Blunt2,Pezeshki-IT08}\cite{Chi-book}. However, as the number of degrees of freedom for transmission increases so does the complexity of the waveform design problem. This motivates the assembly of full waveforms from a  library with simple component waveforms. By choosing to separate waveforms across space, time, frequency, polarization, wavenumber, or a combination of these, we can modularize the design problem.

In this paper, we consider a waveform library consisting of simply two component waveforms. We show that by properly sequencing these component waveforms across time we can construct transmit pulse trains and receive filters for which the radar point-spread function, given by the cross-ambiguity function of the transmit pulse train and the pulse train used in the receive filter, is essentially free of range sidelobes inside an  interval around the zero-Doppler axis. This enables us to extract a weak target that is located in range near a stronger reflector at a different Doppler frequency.


The component waveforms are Golay complementary and are obtained by phase coding a narrow pulse with a pair of Golay complementary sequences (see, e.g., \cite{Golay-IRE61}--\nocite{Welti-IT60}\cite{Tseng-IT72}). Golay complementary sequences have the property that the sum of their autocorrelation functions vanishes at all nonzero lags. Consequently, if the waveforms phase coded by complementary sequences are transmitted separately in time and their ambiguity functions are added together the sum of the ambiguity functions will be essentially an impulse in range along the zero-Doppler axis. This makes Golay complementary waveforms ideal for separating point targets in range when the targets have the same Doppler frequency. However, off the zero-Doppler axis the impulse-like response in range is not maintained and the sum of the ambiguity functions has range sidelobes. In consequence, a weak target that is located in range near a strong reflector with a different Doppler frequency may be masked by the range sidelobes of the radar ambiguity function centered at the delay-Doppler position of the stronger reflector.

We show in this paper that by properly designing the way Golay complementary waveforms are assembled across time in the transmit pulse train and the receive filter, we can essentially annihilate range sidelobes of the radar point-spread function and maintain an impulse-like point-spread function in range over a Doppler interval around the zero-Doppler axis. We construct the transmit pulse train by coordinating the transmission of Golay complementary waveforms according to zeros and ones in a binary sequence $\mathcal{P}$. We refer to this pulse train as the $\mathcal{P}$-pulse train. The pulse train used in the receive filter is constructed in a similar way, in terms of sequencing the Golay waveforms, but each waveform in the pulse train is weighted according to an element of a sequence $\mathcal{Q}$. We call this pulse train the $\mathcal{Q}$-pulse train. The cross-ambiguity function of the $\mathcal{P}$- and $\mathcal{Q}$-pulse trains gives the radar point-spread function, whose shape determines our ability to detect point targets in range and Doppler. We show that the size of the range sidelobes of this cross-ambiguity function is controlled by the spectrum of the product of $\mathcal{P}$ and $\mathcal{Q}$ sequences. By selecting sequences for which the spectrum of their product has a higher-order null around zero Doppler, we can annihilate the range sidelobe of the cross ambiguity function inside an interval around the zero-Doppler axis. However, the signal-to-noise ratio (SNR) at the receiver output, defined as the ratio of the peak of the squared cross-ambiguity function to the noise power at the receiver output, depends on the choice of $\mathcal{Q}$. By jointly designing the transmit-receive sequences $(\mathcal{P},\mathcal{Q})$, we can achieve a trade-off between the order of the spectral null and the output SNR.

We first present two specific $(\mathcal{P},\mathcal{Q})$ designs, namely the \textit{PTM design} and the \textit{Binomial design}, corresponding to the two ends of the trade-off. In the former, the transmit sequence $\mathcal{P}$ is the so-called Prouhet-Thue-Morse (PTM) sequence (see, e.g., \cite{Allouche-SETA98}) of length $N$ and the weighting sequence $\mathcal{Q}$ at the receiver is the all one sequence. In this case, the output SNR in white noise is maximum, as the receiver filter is in fact a matched filter. However, the order of the spectral null is only logarithmic in the length $N$ of the transmit pulse train. In the latter design, $\mathcal{P}$ is the alternating binary sequence of length $N$ and $\mathcal{Q}$ is the sequence of binomial coefficients in the binomial expansion $(1+x)^{N-1}$. In this case, the order of the spectral null is $N-2$, which is the largest that can be achieved with a pulse train of length $N$. However, this comes at the expense of SNR.

We then establish a necessary and sufficient condition for achieving an $M$th-order spectral null with length-$N$, $N>M+1$, sequences $\mathcal{P}$ and  $\mathcal{Q}$. The condition is that the product of the $\mathcal{P}$ and $\mathcal{Q}$ sequences must be in the null space of an $(M+1)\times N$ integer Vandermonde matrix, whose $(m,n)$th element is $(n+1)^m$ for $m=0,1,\ldots,M$ and $n=0,1,\ldots,N-1$. Without additional constraints, there are infinite number of solutions to the problem. In this paper, we constrain $\mathcal{Q}$ to be a positive integer sequence, though other designs are certainly possible. Given a pulse train of length $N$ and a desired null of order $M$, we can then maximize the output SNR to determine a solution for $\mathcal{P}$ and $\mathcal{Q}$.

The PTM design was originally proposed in our earlier papers \cite{Pezeshki-IT08, Chi-book} for constructing Doppler resilient pulse trains of Golay complementary waveforms. This paper extends our previous work to the \textit{joint} design of transmit pulse trains and receive filters. The paper is  intended to provide a summary of results. Proofs have been omitted for brevity.

\section{$(\mathcal{P}$,$\mathcal{Q})$ Pulse Trains}

\textit{Definition 1.} \cite{Golay-IRE61}--\nocite{Welti-IT60}\cite{Tseng-IT72} Two length $L$ unimodular sequences of complex numbers $x(\ell)$ and
$y(\ell)$ are \textit{Golay complementary} if for $k=-(L-1),\ldots,(L-1)$
the sum of their autocorrelation functions satisfies
\[
C_x(k)+C_y(k)=2L\delta(k),
\]
where $C_x(k)$ is the autocorrelation of $x(\ell)$ at lag $k$ and
$\delta(k)$ is the Kronecker delta function.

Consider a pair of waveforms $x(t)$ and $y(t)$ that are phase
coded by length-$L$ complementary sequences $x(\ell)$ and
$y(\ell)$: that is,
\begin{equation}\label{eq:phc}
x(t)=\sum\limits_{\ell=0}^{L-1}x(\ell)s(t-\ell t_c) \ \ 
\text{and} \ \ y(t)=\sum\limits_{\ell=0}^{L-1}y(\ell)s(t-\ell
t_c) \quad
\end{equation}
where $s(t)$ is a baseband pulse shape with unit energy and duration limited to a
chip interval $t_c$.

\textit{Definition 2.} Let $\mathcal{P}=\{p_n\}_{n=0}^{N-1}$ be a discrete binary
sequence of length $N$. Define the \textit{$\mathcal{P}$-pulse train} 
$z_\mathcal{P}(t)$ as
\begin{equation}\label{eq:ptmt}
z_{\mathcal{P}}(t)=\sum\limits_{n=0}^{N-1} p_n
x(t-nT)+\overline{p}_n y(t-nT)
\end{equation}
where $\overline{p}_n=1-p_n$ is the complement of $p_n$. The $n$th
entry in the pulse train is $x(t)$ if $p_n=1$ and is $y(t)$ if
$p_n=0$. Consecutive entries in the pulse train are separated in
time by a PRI $T$.

\textit{Definition 3.} Let
$\mathcal{Q}=\{q_n\}_{n=0}^{N-1}$ be a discrete sequence of length $N$, with positive values $q_n>0$. Define the \textit{$\mathcal{Q}$-pulse train} $z_\mathcal{Q}(t)$ as
\begin{equation}
z_{\mathcal{Q}}(t)=\sum_{n=0}^{N-1}q_{n}\left[p_n x(t-nT)+\overline{p}_n y(t-nT)\right].
\end{equation}
The $n$th element of $z_{\mathcal{Q}}(t)$ is obtained by multiplying the $n$th element of $z_{\mathcal{P}}(t)$ by $q_n$.

If $z_{\mathcal{P}}(t)$ is transmitted by the radar and the return is filtered (correlated) with $z_{\mathcal{Q}}(t)$, then the receiver point-spread function in delay and Doppler will be the cross-ambiguity function between $z_{\mathcal{P}}(t)$ and $z_{\mathcal{Q}}(t)$. After discretizing in delay (at chip intervals), and ignoring the Doppler shift over chip intervals compared to the Doppler shift across a PRI, this cross-ambiguity function is given by
\setlength\arraycolsep{0pt}
\begin{align}\label{eq:ptmamb}
\chi_{\mathcal{P}\mathcal{Q}}\left(k,\theta\right)&=\cfrac{1}{2}\left[C_x(k)+C_y(k)\right]\sum\limits_{n=0}^{N-1}q_{n}e^{jn\theta}\nonumber\\
&+\cfrac{1}{2}\left[C_x(k)-C_y(k)\right]\sum\limits_{n=0}^{N-1}(-1)^{p_{n}}q_{n}
e^{jn\theta}
\end{align}
where $\theta$ is the relative Doppler shift over a PRI $T$. Since $x(k)$ and $y(k)$ are Golay complementary,  $C_x(k)+C_y(k)=2L\delta(k)$
and the first term on the right-hand-side of \eqref{eq:ptmamb} is
free of range sidelobes. The second term represents the range
sidelobes, as $C_x(k)-C_y(k)$ does not vanish at all $k\neq 0$.

\textit{Controlling Range Sidelobes.} The magnitude of the range sidelobes is proportional to the magnitude of the spectrum of the sequence $(-1)^{p_n}q_n$, given by
\begin{equation}\label{eq:Sp}
S_{\mathcal{P},\mathcal{Q}}(\theta)=\sum_{n=0}^{N-1}(-1)^{p_{n}}q_{n}e^{jn\theta}.
\end{equation}
As a result, range sidelobes inside a Doppler interval around the zero-Doppler axis can be suppressed by selecting a sequence $(-1)^{p_n}q_n$ whose spectrum has a higher-order null at zero Doppler.

Consider the Taylor expansion of $S_{\mathcal{P},\mathcal{Q}}(\theta)$
around $\theta=0$, that is,  
\begin{equation}
S_{\mathcal{P},\mathcal{Q}}(\theta)=\sum_{m=0}^{\infty}S_{\mathcal{P},\mathcal{Q}}^{(m)}(0)\frac{\theta^m}{m!}
\end{equation}
where $S_{\mathcal{P},\mathcal{Q}}^{(m)}(0)$ is the $m$-th order derivative of $S_{\mathcal{P},\mathcal{Q}}(\theta)$ at $\theta=0$. To create an $M$th order null, all $S_{\mathcal{P},\mathcal{Q}}^{(m)}(0)$ up to order $M$ must vanish: that is, 
\begin{equation}\label{eq:null_space}
S_{\mathcal{P},\mathcal{Q}}^{(m)}(0)=0,\ m = 0,1,...,M,
\end{equation}
or equivalently, 
\begin{equation}\label{eq:null_space1}
\sum_{n=0}^{N-1}n^m(-1)^{p_n}q_n = 0,\ m=0,1,...,M.
\end{equation}

\textit{Controlling Signal-to-Noise Ratio.} Suppose the noise at the receiver input is white and has power $N_0$. Then, the noise power at the receiver output is
\begin{equation}
\eta=N_0\int_{-\infty}^{\infty}|z_{\mathcal{Q}}(t)|^2dt =N_0L\|\mathbf{q}\|_2^2,
\end{equation}
where $\mathbf{q}=[q_0,...,q_{N-1}]^T$. The SNR at the receiver output is given by 
\begin{equation}
\rho = \frac{\sigma_b^2|\chi_{\mathcal{P},\mathcal{Q}}(0,0)|^2}{\eta}=\frac{L\sigma_b^2}
{N_0}\frac{\|\mathbf{q}\|_1^2}{\|\mathbf{q}\|_2^2},
\end{equation}
where $\sigma_b^2$ is the variance of the scattering coefficient of the target.

The SNR $\rho$ is maximized when $\mathbf{q}=\alpha\mathbf{1}$ for some positive scalar $\alpha$, meaning that $z_{\mathcal{Q}}(t)=\alpha z_{\mathcal{P}}(t)$ which corresponds to the usual matched filter. Any sequence $\mathcal{Q}$ other than the all one sequence results in a reduction in SNR. However, the extra degrees of freedom provided by a more general $\mathcal{Q}$ can be used to create a spectral null of higher order, through the joint design of $\mathcal{P}$ and $\mathcal{Q}$, than what is achievable by only designing $\mathcal{P}$.

\textit{Design Trade-off.} The joint design of $\mathcal{P}$ and $\mathcal{Q}$ sequences enables a trade-off between the order of the spectral null for range sidelobe suppression around zero Doppler and the SNR at the receiver output. In the next section, we first present two examples of $(\mathcal{P},\mathcal{Q})$ designs, namely the \emph{PTM design} (see also \cite{Pezeshki-IT08, Chi-book}) for which the order of the spectral null is logarithmic in the pulse train length $N$, and the \emph{Binomial design} for which the order of the null is linear in $N$. The latter design maintains an impulse-like point-spread function in range over a wider Doppler interval around the zero-Doppler axis. But this added invariance comes at the expense of SNR. Later, we derive necessary and sufficient conditions for achieving an $M$th order spectral null with a pulse train of length $N$ and further investigate the trade-off.

\begin{figure*}\label{f:PTM}
\begin{center}
\begin{tabular}{ccc}
\subfigure[]
{\includegraphics[width=3.2in]{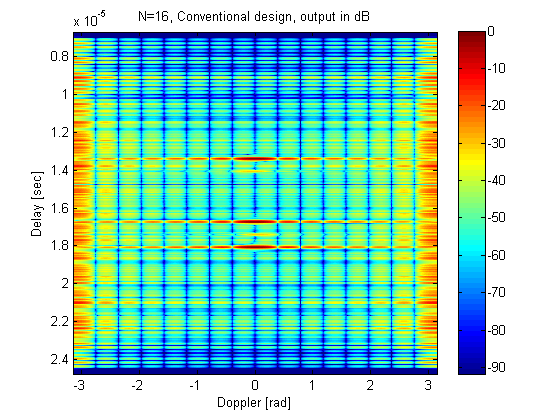}}&
\subfigure[]
{\includegraphics[width=3.2in]{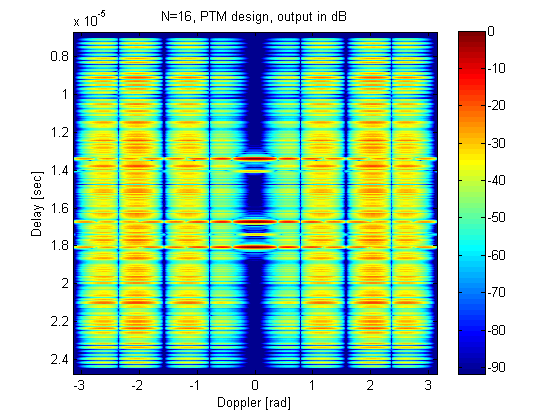}}\\
\subfigure[]
{\includegraphics[width=3.2in]{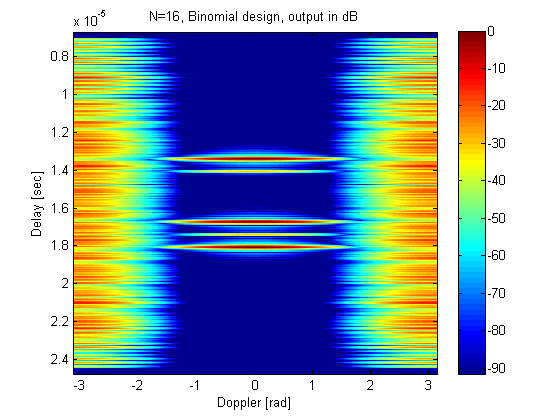}}&
\subfigure[]
{\includegraphics[width=3.2in]{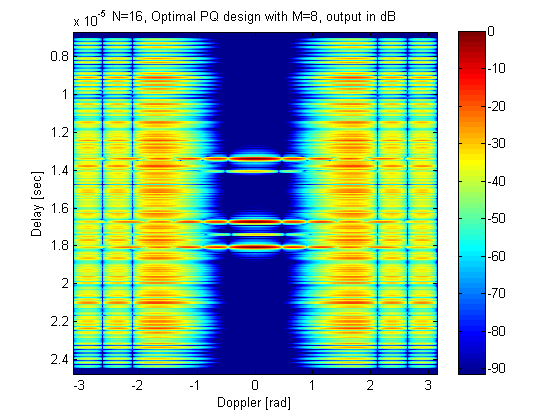}}
\end{tabular}
\end{center}
\caption{Comparison of output delay-Doppler maps for different $(\mathcal{P},\mathcal{Q})$ designs: (a) conventional design, (b) PTM design, (c) Binomial design, and (d)  max-SNR design with an 8-th order null. The scene contains three strong (equal amplitude) stationary reflectors at different ranges and two weak slow moving targets (30dB weaker).}
\end{figure*}

\section{Range Sidelobe Suppression}

\textit{Theorem 1: PTM Design.} Let $\mathcal{P}=\{p_n\}_{n=0}^{N-1}$ be the length $N=2^{M+1}$ Prouhet-Thue-Morse (PTM) sequence (see, e.g., \cite{Allouche-SETA98}), defined recursively as $p_{2k}=p_{k}$ and $p_{2k+1}=1-p_{k}$ for all $k\geq 0$, with $p_0=0$, and let $\mathcal{Q}=\{q_n\}_{n=0}^{N-1}$ be the all 1 sequence of length $N=2^{M+1}$. Then,  $S_{\mathcal{PQ}}(\theta)$ has an $M$th-order null at $\theta=0$.

\textit{Example 1.} The PTM sequence of length $N=4$ is $\mathcal{P}=(p_k)_{k=0}^{3} \ = \ 0 \ 1 \ 1 \ 0$. The corresponding $\mathcal{P}$-pulse train of Golay complementary waveforms is given by
\begin{align}
z_{\mathcal{P}}(t)= x(t)+y(t-T)+y(t-2T)+x(t-3T).\nonumber
\end{align}
The receive filter pulse train $z_{\mathcal{Q}}(t)$ is the same as the $\mathcal{P}$-pulse train. The order of the spectral null for range sidelobe suppression is $M=\left(\log_{2}N\right)-1=1$.

\textit{Theorem 2: Binomial Design.} Let $\mathcal{P}=\{p_n\}_{n=0}^{N-1}$ be the length $N=M+2$ alternating sequence, where $p_{2k}=1$ and $p_{2k+1}=0$ for all $k\ge 0$, and let $\mathcal{Q}=\{q_n\}_{n=0}^{N-1}$ be the length $N=M+2$ binomial sequence $\{q_{n}\}_{n=0}^{N-1}=\{\binom{N-1}{n}\}_{n=0}^{N-1}$. Then, $S_{\mathcal{PQ}}(\theta)$ has an $M$th order null at $\theta=0$.

\textit{Example 2.} For $N=4$, the $\mathcal{P}$-pulse train transmitted by the radar is
\begin{equation}
z_{\mathcal{P}}(t)=x(t)+y(t-T)+x(t-2T)+y(t-3T) \nonumber
\end{equation}
and the $\mathcal{Q}$-pulse train (binomial) used for filtering is
\begin{align}
z_{\mathcal{Q}}(t)=q_0x(t)+q_1 y(t-T)\quad\quad\quad\nonumber\\
\quad\quad\quad+q_2 x(t-2T)+q_3 y(t-3T)\nonumber
\end{align}
where $q_n=\binom{3}{n}$, $n=0,1,2,3$. The order of the spectral null for sidelobe suppression is $M=N-2=2$.

We now give the general condition for achieving an $M$th-order spectral null with $\mathcal{P}$ and $\mathcal{Q}$ sequences of length $N>M+1$. 

\textit{Theorem 3.} The spectrum $S_{\mathcal{P},\mathcal{Q}}(\theta)$ has an
$M$-th order null, $M<N-1$, at $\theta=0$ if and only if  
\begin{equation}\label{eq:mat_form}
\begin{bmatrix}1\ & 1\ & \cdots & 1 \\
1\ & 2\ & \cdots &N \\
\vdots & \vdots & \ddots & \vdots \\
1^{M}\ & 2^{M}\ & \cdots &N^{M}
\end{bmatrix}\begin{bmatrix}(-1)^{p_0}q_0 \\ (-1)^{p_1}q_1 \\ \vdots
\\(-1)^{p_{N-1}}q_{N-1}\end{bmatrix} = \mathbf{0}.
\end{equation}

\textit{Remark 1.} To avoid trivial solutions, $M$ has to be less than $N-1$. For a given pulse train length $N$, the Binomial design achieves the maximum order $M=N-2$ of spectral null.

\textit{Remark 2.} Let $T(M')$ denote the set of product sequences $\{(-1)^{p_n}q_n\}_{0}^{N-1}$ that satisfy the null space condition (\ref{eq:mat_form}) for $M=M'$. Then, clearly, we have $T(0)\supseteq T(1) \supseteq \cdots \supseteq T(N-2)$. 

Fig. 1 illustrates the annihilation of range-sidelobes around the zero-Doppler axsis for three different length-$16$ $(\mathcal{P},\mathcal{Q})$  designs and compares their delay-Doppler responses with that of a conventional design. The conventional design uses an alternating transmission of Golay complementary waveforms followed by matched filtering at the receiver. The scene contains three strong reflectors of equal amplitudes at different ranges and two weak targets (each 30dB weaker) that have small Doppler frequencies relative to the stronger reflectors. The horizontal axis depicts Doppler and the vertical axis illustrates delay. Color bar values are in dB.

In the conventional design, shown in Fig. 1(a), the weak targets are almost completely masked by the range sidelobes of the stronger reflectors. With the PTM design, shown in Fig. 1(b), we can clear out the range sidelobes inside a narrow Doppler interval around the zero-Doppler axis. The order of the spectral null for range sidelobe suppression in this case is $M=\left(\log_2^{16}\right)-1=3$. With this order, we can bring the range sidelobes below -80dB inside the $[-0.1,-0.1]$ rad Doppler interval and extract the weak targets. If the difference in the Doppler frequencies of the weak and strong reflectors is larger, we need a null of higher order to annihilate the range sidelobes inside a wider Doppler band. Fig. 1(c) shows that the Binomial design (of length $N=16$) can expand the cleared (below -80dB) region to $[-1,-1]$ rad by creating a null of order $M=16-2=14$ around zero Doppler. However, this increase in the order of the spectral null comes at the expense of SNR. Fig. 1(d) shows the delay-Doppler response of a $(\mathcal{P},\mathcal{Q})$ design that has the largest SNR among all $(\mathcal{P},\mathcal{Q})$ designs that achieve an $(M=8)$th order spectral null. The cleared region in this case is the $[-0.5,0.5]$ rad Doppler interval.

Table I compares the three designs in terms of the null order and the output SNR, and shows that by jointly designing the $\mathcal{P}$ and $\mathcal{Q}$ sequences we can achieve a null of relatively high order without considerable reduction in SNR compared to a matched filter design.

\begin{table}[h]
\caption{Null order \& SNR for different designs}
\begin{center}
\begin{tabular}{|c|c|c|}
  \hline
  $(\mathcal{P},\mathcal{Q})$ design & Null order & SNR ($\|\mathbf{q}\|_1^2/\|\mathbf{q}\|_2^2$)\\ \hline
  Conventional & 0 & 16\\ \hline
  PTM  & 3 & 16\\ \hline
  Max-SNR with $M=8$ & 8 & 13.76\\ \hline
  Binomial & 14 & 6.92\\ \hline
\end{tabular}
\end{center}
\end{table}

\section{Conclusion}

In this paper, we have presented a general method for constructing radar transmit pulse trains and receive filters for which the radar point-spread function is essentially free of range sidelobes inside a Doppler interval around the zero-Doppler axis. The radar point-spread function is given by the cross-ambiguity function of the transmitted pulse train and the pulse train used for filtering at the receiver. Controlling  range sidelobes around the zero-Doppler axis enables us to extract weak targets that are located in range near stronger reflectors.

We construct the transmit pulse train and the receive filter by assembling waveforms from a small library of complementary component waveforms. This modular approach allows for managing the design complexity. We employ a binary sequence to coordinate the transmission of a pair of Golay complementary waveforms, with successive transmissions being separated in time by pulse repetition intervals. We build the pulse train used at the receive filter in a similar way in terms of sequencing the complementary waveforms but we employ another sequence to give weights to each term at the receiver. The magnitude of the range sidelobes of the corresponding cross-ambiguity function is controlled by the spectrum of the product of the two sequences. By choosing sequences whose products have higher-order spectral nulls at zero frequency, we can annihilate the range sidelobes of the cross-ambiguity function and maintain an essentially impulse-like point-spread function in range inside an interval around the zero-Doppler axis. The joint design of the two sequences enables a trade-off between the order of the spectral null for range sidelobe suppression and SNR at the receiver output.

\bibliographystyle{IEEEtran}
\bibliography{MIMORadar,DoppRef,linalgebbib}

\end{document}